\documentstyle[12pt]{article}
\input{psfig.sty}
\title{\bf High and low temperature behavior of a quantum group
fermion gas}
\author {\\ \\ \\ Marcelo R. Ubriaco\thanks{ubriaco@ltp.upr.clu.edu}\\
{\em Laboratory of Theoretical Physics}\\
{\em Department of Physics}\\
{\em  University of Puerto Rico}\\
{\em P. O. Box 23343, R\'{\i}o Piedras}\\
{\em PR 00931-3343, USA}}
\date{}
\begin{document}
\maketitle
\vspace*{-.3in}
\begin{abstract}
We consider the simplest  $SU_{q}(2)$ invariant fermionic hamiltonian
and calculate the low and high temperature behavior for the two
distinct cases $q>1$ and $q<1$. For low temperatures we find that 
entropy values  for the Fermi case are an upper bound for those
corresponding to $q\neq 1$.  At high temperatures we find that
the sign of the second virial coefficient depends on  $q$, and vanishes at $q=1.96$.
 An important consequence of this
fact is that the parameter $q$
connects the fermionic and bosonic regions, showing therefore 
that 
$SU_{q}(2)$ fermions exhibit fractional statistics in
three spatial dimensions.
\end{abstract}
\newpage
\baselineskip20pt

The relevance that quantum groups and quantum algebras \cite{Jimbo} may have
outside the scope  of integrable models has been in the last few years 
a subject of much interest for both physicists and mathematicians.  The motivation
of introducing quantum group symmetries in either space-time or as  internal
degrees of freedom addressed new interesting approaches which could give a new
insight on the diverse roles they may play in physics.  
Some of these approaches led to
the formulation of noncommutative geometry \cite{Wo,Manin,WZ},
and numerous studies in  quantum mechanics
\cite{Wat}, field theory \cite{AV}, molecular and nuclear physics
\cite{I}.  The main objective in this article is to investigate 
 the thermodynamics of a gas with constituents
satisfying an algebra covariant under
the group $SU_q(N)$. The thermodynamics of $q$-deformed systems \cite{Th}
and their possible relevance
to anyon statistics \cite{An}
has been studied by several authors exclusively in the context
of the so called  quons \cite{G}.
In our case,  we consider as our
starting point
 a very simple quantum group invariant hamiltonian.  As the most
natural choice we consider a free $SU_{q}(2)$ invariant hamiltonian 
in terms of operators generating a $SU_{q}(2)$ covariant algebra, which 
in the $q\rightarrow 1$ limit will become a fermionic gas.   
We study this model for low (high density) and high (low density) temperatures.  
At low $T$ we find that the entropy function
 for quantum group
gases, $q\neq 1$, lies below the graph of the fermionic gas entropy. At a 
given temperature the lowest entropy values corresponds to the extreme case of
$q\rightarrow 0$.  At high $T$
we find that the second virial coefficient as 
a function of the parameter $q$ interpolates continuously from fermionic to bosonic
behavior, and it vanishes at $q=1.96$.

The $SU_{q}(2)$ covariant algebra generated by
the quantum  group fermions $\Psi_{i}$, $i=1,2$, is given
by the following relations
\begin{equation}
\{\Psi_{2},\overline{\Psi}_{2}\}=1 \label{1}
\end{equation}
\begin{equation}
\{\Psi_{1},\overline{\Psi}_{1}\}=1 - (1-q^{-2})\overline{\Psi}_{2}\Psi_{2}
\end{equation}
\begin{equation} 
\Psi_{1}\Psi_{2}=-q \Psi_{2}\Psi_{1}
\end{equation}
\begin{equation} 
\overline{\Psi}_{1}\Psi_{2}=-q \Psi_{2}\overline{\Psi}_{1}
\end{equation}
\begin{equation}
\{\Psi_{1},\Psi_{1}\}=0=\{\Psi_{2},\Psi_{2}\} \label{0},
\end{equation}
which for $q=1$ become a $SU(2)$ covariant  fermionic algebra. 
The covariance is simply checked by applying  the linear
transformation $\Psi'_{i}=\sum_{j=1}^{2}T_{ij}\Psi_{j}$, where
the matrix $T=\left(\begin{array}{cc} a & b \\ c & d\end{array}\right)$  is 
the two dimensional representation of the quantum group  $SU_{q}(2)$ 
\cite{T} and the matrix coefficients $(a,b,c,d)$ generate the algebra
\begin{eqnarray}
ab=q^{-1}ba  & , & ac=q^{-1}ca \nonumber \\
bc=cb & , & dc=qcd  \nonumber \\
db=qbd & , &  da-ad=(q-q^{-1})bc  \nonumber \\
& & det_{q}T\equiv ad-q^{-1}bc=1  .
\end{eqnarray}
Requiring  $T$ to be unitary leads to the adjoint matrix 
$\overline{T}$ given by
\begin{equation}
\overline{T}=\left(\begin{array}{cc} d & -qb \\ -q^{-1}c & a\end{array}\right),
\end{equation}
where the parameter $q$ must be a real number.  Hereafter, 
we will consider $0\leq q<\infty$. In general, $SU_{q}(N)$ covariant
fermionic algebras can be written in terms of the $R$-matrix of
$\hat{A}^{q}_{N-1}$, as given in Reference  \cite{U2}.

There is a clear distinction between the algebra in Equations (\ref{1})-(\ref{0})
with the $q$-boson algebra. Quons $a_i$ obey the relations
\begin{equation}
a_ia_j^{\dagger}-qa_{j}^{\dagger}a_i=\delta_{i,j},\label{quon}
\end{equation}
which has the boson and fermion limits for $q=1$ and $q=-1$ respectively.
No specific relation between annihilation (or creation) operators
is known for $q^2\neq 1$ and
Equation (\ref{quon}) is not covariant under $SU_q(N)$.

A representation of $SU_{q}(N)$-fermions in terms of fermionic operators 
$\psi_{i}$
and $\psi_{j}^{\dagger}$  is given by the relations
\begin{equation}
\Psi_{m}=\psi_{m}\prod_{l=m+1}^{N}\left(1+(q^{-1}-1) M_{l}\right)\label{psi},
\end{equation}
\begin{equation}
\overline{\Psi}_{m}=\psi_{m}^{\dagger}\prod_{l=m+1}^{N}
\left(1+(q^{-1}-1) M_{l}\right)\label{psi1},
\end{equation}
where $M_{l}=\psi_{l}^{\dagger}\psi_{l}$ and $\{\psi_{i},\psi_{j}^{\dagger}\}=
\delta_{ij}$.
In what follows we consider the simplest $SU_{q}(2)$ invariant hamiltonian,
 which is
\begin{equation}
{\cal H}=\sum_{\kappa}^{}\varepsilon_{\kappa}(\overline{\Psi}_{\kappa,1}{\Psi}_{\kappa,1}
+\overline{\Psi}_{\kappa,2}{\Psi}_{\kappa,2}), \label{H}
\end{equation}
with the quantum group fields satisfying $\{\overline{\Psi}_{\kappa,i},
\Psi_{\kappa',j}\}=0$ for $\kappa\neq\kappa'$. Thus, in terms of the fermionic fields
$\psi_{i}$, Equation (\ref{H}) becomes the interacting hamiltonian 
\begin{equation}
{\cal H}=\sum_{\kappa}^{}\varepsilon_{\kappa}(M_{\kappa,1}+
M_{\kappa,2}+(q^{-2}-1) M_{\kappa,1}M_{\kappa,2}),
\end{equation}
where $M_{\kappa,i}
=\overline{\psi}_{\kappa,i}\psi_{\kappa,i}$ is the fermionic number operator .  
The grand partition function
${\cal Z}$ is 
\begin{equation}
{\cal Z}=\prod_{\kappa}(1+2e^{-\beta (\varepsilon_{\kappa}-\mu)}+
e^{-\beta (\varepsilon_{\kappa}(q^{-2}+1)-2\mu)})\label{z},
\end{equation}
which at $q=1$ becomes the partition function of a fermion gas 
with two degrees of freedom.
The average number of particles $\langle M\rangle$ becomes then a function of the parameter
$q$ according to the equation
\begin{eqnarray}
\langle M\rangle&=&\langle m_{1}\rangle +\langle  m_{2}\rangle 
\nonumber \\
&=&2V\int_{}^{}\frac{e^{\beta(\mu-\frac{p^{2}}{2m})}
(1+e^{\beta(\mu-\frac{q^{-2}p^{2}}{2m}})d^{3}p}{(2\pi\hbar)^{3}
(1+2e^{-\beta (\frac{p^{2}}{2m}-\mu)}+
e^{-\beta (\frac{p^{2}}{2m}(q^{-2}+1)-2\mu)})}.\label{M}
\end{eqnarray}
\\
{\em Low temperature and high density gas}

In a previous work \cite{U2} we studied Equation (\ref{M}) for the 
extreme cases $q\gg 1$ and $q\ll 1$.  It was shown that for low
temperatures  the chemical potential  $\mu(T)$
for  $q\gg 1$ is almost identical to the
Fermi case while that for $q\ll 1$ this function has in addition a 
linear temperature dependent term which vanishes for $q=1$ and
$q\gg 1$.  
In this section we calculate the function $\langle M\rangle$ for the two cases
$q>1$ and $q<1$. In addition, 
we calculate the internal energy
$U$, heat capacity $C_{v}$ and entropy $S$ and compare them with the
fermionic, $q=1$, case.
\begin{description}
\item[a)] $q>1$\\
A simple way to calculate
Equation (\ref{M}) is by considering that at $T=0$ the 
average number of particles is given by
$\langle M\rangle=\frac{2}{3(q^{-2}+1)^{3/2}}\lambda(2\mu_{0})^{3/2}$. Thus, we split this
equation in the
following integrals
\begin{eqnarray}
\langle M\rangle&=&\lambda\int_{0}^{2\mu/(q^{-2}+1)}\varepsilon^{1/2}d\varepsilon
\nonumber \\ 
&-&\lambda\left(\int_{0}^{\mu}+\int_{\mu}^{2\mu/(q^{-2}+1)}\right)
\varepsilon^{1/2}
\frac{1+e^{\beta(\varepsilon-\mu)}}{f(\varepsilon,\mu,q)}d\varepsilon \nonumber\\
&+&\lambda\left(\int_{2\mu/(q^{-2}+1)}^{q^{2}\mu}+
\int_{q^{2}\mu}^{\infty}\right)\varepsilon^{1/2}\frac{1+e^{-\beta(q^{-2}\varepsilon-\mu)}}
{f(\varepsilon,\mu,q)}d\varepsilon,\label{M0}
\end{eqnarray}
where $f(\varepsilon,\mu,q)=e^{\beta(\varepsilon-\mu)}+2
+e^{-\beta(q^{-2}\varepsilon-\mu)}$ and $\lambda=\frac{4\pi V(2m)^{3/2}}{(2\pi\hbar)^{3}}$. 
The second and fifth integrations vanish as
$\beta\mu\rightarrow \infty$, and the remaining terms lead to the result
\begin{equation}
\langle M\rangle\approx\frac{2}{3(q^{-2}+1)^{3/2}}\lambda(2\mu)^{3/2}+\frac{0.64\lambda}
{(q^{-2}+1)^{3/2}\beta^{2}\sqrt{2\mu}}.\label{M1}
\end{equation}
Equation (\ref{M1}) is very similar to the Fermi case in the
context that it does not contain a linear term in $T$.
 The internal energy $U$ is calculated from
the grand potential $\Omega=-\frac{1}{\beta}\ln{\cal Z}$ 
\begin{eqnarray}
U&=&(\frac{\partial\beta\Omega}{\partial\beta}+\mu M)\nonumber\\
&=&V\int_{}^{}\frac{p^{2}}{2m}\frac{
(2+(q^{-2}+1)e^{\beta(\mu-\frac{q^{-2}p^{2}}{2m}})d^{3}p}{(2\pi\hbar)^{3}
f(\varepsilon,\mu,q)},\label{U}
\end{eqnarray}
such that following the same procedure as in Equation (\ref{M0})
we obtain for the internal energy  
 \begin{equation}
U\approx \lambda\left[\left(\frac{2}{q^{-2}+1}\right)^{3/2}\frac{2}{5}
\mu_{0}^{5/2}+\frac{0.64\sqrt{2\mu_{0}}}{(q^{-2}+1)^{3/2}\beta^{2}}\right],
\end{equation}
and the entropy
\begin{equation}
S\approx\lambda\frac{1.28\sqrt{2\mu_{0}}k^{2}T}{(q^{-2}+1)^{3/2}}.
\end{equation}
\item[b)] $q<1$\\
For $q<1$ the function $\langle M\rangle$ was calculated in
Reference \cite{U2}. We use this result 
as a starting point and then calculate some thermodynamic
functions.  The average number of particles 
is given by the equation
\begin{equation}
\langle M\rangle\approx\frac{\lambda\mu^{3/2}(1+q^{3})}{3}+\frac{\lambda 
\sqrt{\mu}(1-q^{3})\ln 3}{2\beta}
+\frac{ 0.54 \lambda (1+q^{3})}{2\sqrt{\mu}\beta^{2}},\label{M2}
\end{equation}
which contains a linear temperature term that vanishes for
$q=1$. Reverting Equation (\ref{M2}) leads to the chemical potential
function
\begin{equation}
\mu\approx\mu_{0}\left[1-\frac{(1-q^{3})\ln3}{(1+q^{3})\mu_{0}}kT
-\left(0.54-\frac{(1-q^{3})^{2}\ln^{2}3}{4(1+q^{3})^{2}}\right)\left(\frac{kT}{\mu_{0}}\right)^{2}\right]
\label{mu}.
\end{equation}
The internal energy $U$ is simply  calculated from Equation (\ref{U})
 by splitting this integral in 
the intervals $[0,q^{2}\mu]$,
$[q^{2}\mu,\mu]$ and $[\mu,\infty]$.  The internal energy
in terms of the chemical potential is given by the equation
\begin{equation}
U\approx\frac{\lambda}{2}\left[\frac{2}{5}(q^{3}+1)\mu^{5/2}
+\frac{\ln3}{\beta}(1-q^{3})\mu^{3/2}+\frac{3}{\beta^{2}}
0.54(1+q^{3})\sqrt{\mu}\right],
\end{equation}
such that after replacement of Equation (\ref{mu}) 
into this equation gives for the internal energy
\begin{equation}
U\approx\frac{\lambda}{2}\left[\frac{2}{5}(q^{3}+1)\mu_{0}^{5/2}+
\left(1.08(q^{3}+1)-\frac{(1-q^{3})^{2}}{2(1+q^{3})}\ln^{2}3\right)
\frac{\mu_{0}^{1/2}}{\beta^{2}}\right].\label{U1}
\end{equation}
The linear term in $T$ has canceled out and the heat
capacity
\begin{eqnarray} 
C_{v}&=&\left(\frac{\partial U}{\partial T}\right)_{V}\nonumber\\
C_{v}&\approx&\lambda k^{2}\sqrt{\mu_{0}}T\left[1.08(q^{3}+1)
-\frac{(1-q^{3})^{2}}{2(1+q^{3})}\ln^{2}3\right],
\end{eqnarray}
 vanishes at $T=0$ in accordance with the third law. There is no
 solution with $q\in {\bf R}$ such that $C_{v}=0$, and
there is only  one real solution for $q<1$ at a given
temperature for each value of $C_{v}$.

Figure 1 shows the function $s=S/\lambda k^{2}\sqrt{\mu_{0}}$
as a function of $T$ for different values of the parameter $q$,
where the entropy $S=\int_{0}^{T}\frac{C_{v}}{T'}dT'$. From  Figure 1
we see that the entropy is  maximum for the Fermi case, and there are
systems
with $q>1$ sharing the entropy function with those with $q<1$.  Specifically,
for every value $q_{1}$ such that $0.33\leq q_{1}<0.91$ there is a 
solution $q_{2}>1$ with the same entropy function.  For other values
of $q_{1}$ there is no value $q_{2}>1$ which would give the
same entropy at a given $T$.  Therefore, for low temperatures  entropy functions given
by hamiltonians with $q>1$
are contained in the set of those with $q<1$.
\begin{figure}
\centerline{\psfig{figure=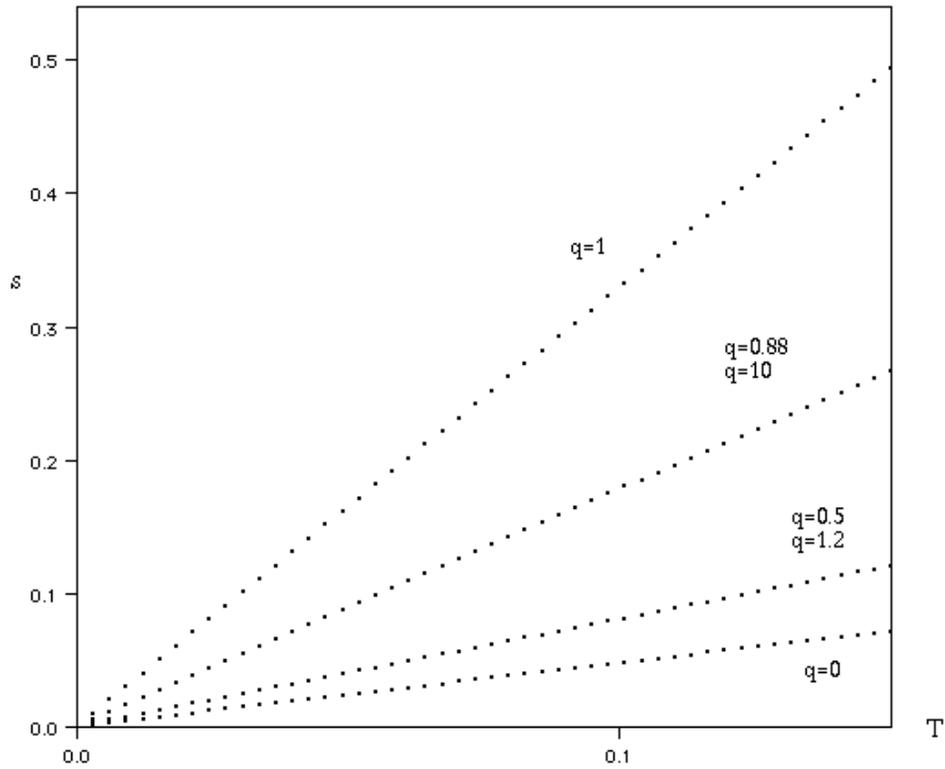,height=5in}}
\caption{The function $s$ as defined in the text for
low temperatures and several values of $q$}
\end{figure}
\end{description}
{\em High temperature and low density gas}\\

In this section we calculate several thermodynamic functions for the case
of weak degeneracy $z=e^{\beta\mu}\ll 1$, and  study the role played by
the parameter $q$ in the equation of state.  Our starting point is the grand
partition function ${\cal Z}$
\begin{equation}
\ln {\cal Z}=\frac{4\pi V}{(2\pi\hbar)^{3}}\int_{0}^{\infty}p^{2}
\ln\left(1+2e^{-\beta(\varepsilon-\mu)}+e^{-\beta(\varepsilon(q^{-2}+1)-2\mu)}\right)dp,
\end{equation}
such that expanding the integrand and keeping the first three terms  gives
\begin{equation}
\ln {\cal Z}=\frac{\lambda\sqrt{\pi}}{\beta^{3/2}}\left[\frac{z}{2}-\alpha(q) \frac{z^{2}}{2}
+\gamma(q)\frac{z^{3}}{3!}+...\right],\label{z1}
\end{equation}
where the functions $\alpha(q)$ and $\gamma(q)$ are
\begin{eqnarray}
\alpha(q)&=&\frac{1}{2^{3/2}}-\frac{1}{2(q^{-2}+1)^{3/2}}\nonumber\\ 
\gamma(q)&=&\frac{4}{3^{3/2}}-\frac{3}{(q^{-2}+2)^{3/2}}.\nonumber
\end{eqnarray}
Once we calculate the average number of particles $\langle M\rangle=\frac{1}{\beta}\left(\frac{\partial\ln{\cal Z}}{\partial\mu}
\right)_{T,V}$ we can write the fugacity $z$ in terms of $\langle M\rangle$, and Equation (\ref{z1})
becomes 
\begin{equation}
\ln{\cal Z}=\langle M\rangle\left[1+\frac{2\alpha(q)\langle M\rangle}{\lambda\sqrt{\pi}}\beta^{3/2}
-\frac{\langle M\rangle^{2}}{\lambda^{2}\pi}\beta^{3}\Lambda+...\right],
\end{equation}
where $\Lambda=\frac{8\gamma(q)}{3}+16\alpha^{2}(q)$.\\
From this equation we can obtain the internal energy $U$, the heat capacity $C_{v}$ and
the entropy $S=\frac{U-\mu\langle M\rangle}{T}+k\ln{\cal Z}$ as functions of $\langle M\rangle$. The corresponding
equations are
\begin{equation}
U=\frac{3\langle M\rangle}{2\beta}+\frac{3\langle M\rangle^{2}\beta^{1/2}\alpha(q)}{\sqrt{\pi}\lambda}
-\frac{3\langle M\rangle^{3}\beta^{2}}{2\pi\lambda^{2}}\Lambda+...,
\end{equation}
\begin{equation}
C_{v}=\frac{3\langle M\rangle k}{2}\left[1-\frac{\langle M\rangle\alpha(q)\beta^{3/2}}{\lambda\sqrt{\pi}}
+\frac{2\langle M\rangle^{2}\beta^{3}}{\lambda^{2}\pi}\Lambda+...\right],
\end{equation}
\begin{figure}
\centerline{\psfig{figure=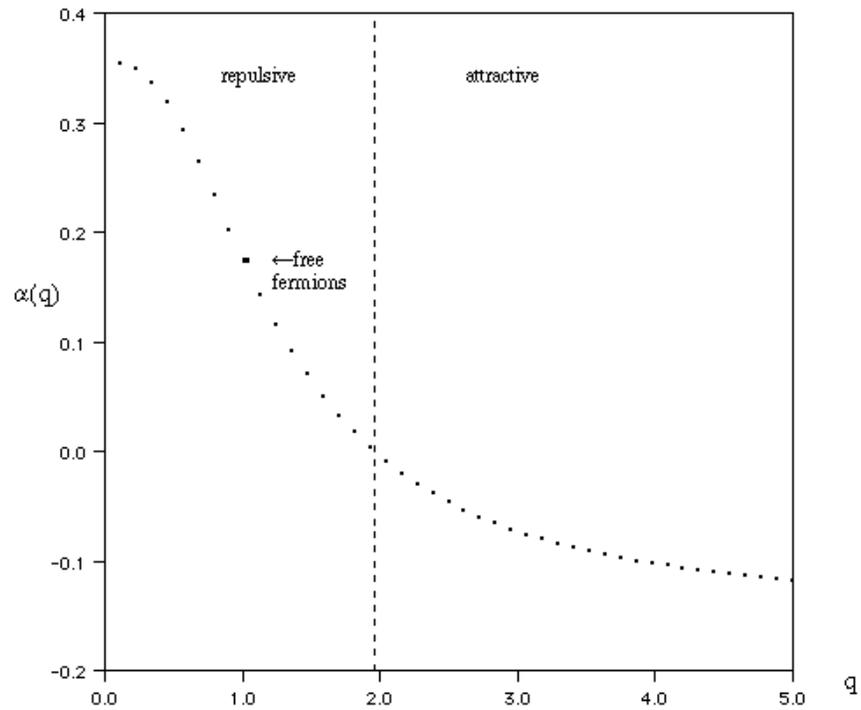,height=5in}}
\caption{The coefficient $\alpha(q)$ for the interval
$0\leq q\leq 5$. The line at $q=1.96$
divides the region between $\alpha(q)>0$ and $\alpha(q)<0$ which corresponds to
fermionic and boson-like behavior respectively}
\end{figure}
\begin{equation}
S=\langle M\rangle k\left[\frac{5}{2}-\ln\left(\frac{2\langle M\rangle\beta^{3/2}}{\lambda\sqrt{\pi}}\right)
+\frac{\alpha(q)\langle M\rangle}{\lambda\sqrt{\pi}}\beta^{3/2}+...\right].
\end{equation}
The equation of state  is given by the equation
\begin{equation}
pV=kT\langle M\rangle\left[1+\frac{2\alpha(q)\beta^{3/2}\langle M\rangle}{\lambda\sqrt{\pi}}+...\right],
\end{equation}
with the second virial coefficient $B_{2}(q)$  
\begin{equation}
B_{2}(q)=\frac{\alpha(q)}{2}\left(\frac{h^{2}}{2mkT\pi}\right)^{3/2}.
\end{equation}
The sign of the second virial coefficient clearly depends of the value of $q$,
therefore the parameter $q$ interpolates between fermion-like and boson-like behavior.
Figure 2 shows a graph of the coefficient $\alpha(q)$ as a function of $q$.
The function $\alpha(q)$ takes
values in the interval $2^{-5/2}\leq\alpha\leq 2^{-3/2}$
for  $0\leq\ q\leq 1$, vanishes at $q=1.96$ and gets its lowest
 value  $\alpha(q)=-0.15$ in the limit $q\rightarrow\infty$.
One question one
could address is whether
performing the same calculation 
 in two dimensions it would lead to
a similar behavior, such that a possible connection between the
anyonic and our second virial coefficient for the two dimensional
system could be established.
Repeating the same procedure for  two
spatial dimensions one finds that the second virial coefficient becomes the function 
\begin{equation}
B_{2}(q)=\frac{\pi\hbar^{2}\beta}{2m(q^{2}+1)},
\end{equation}
 which
is positive for all values of $q$, showing therefore that this 
system does not exhibit anyonic behavior in two dimensions.

In this article we studied the behavior of a $SU_{q}(2)$ fermionic gas
at low and high temperatures. We calculated 
 several thermodynamic functions for the two cases $q>1$ and $q<1$. 
Our results point out that
at a low $T$ the entropy value for a gas with $q\neq 1$ is lower
than the one for the Fermi case, and become the lowest for $q\rightarrow 0$.  
In particular, for  $0.33\leq q< 0.91$ systems 
with $q>1$ share the entropy function with systems
with $q<1$. 
We studied the behavior of this model at high temperatures
and obtained the equation of state as a virial expansion.
We found that  the second virial coefficient $B_{2}(q)$ has a dependence
on the parameter $q$ such that it vanishes at $q=1.96$
and becomes negative for $q>1.96$.  Thus, as the
parameter $q$ varies form zero to infinity, this
simple quantum group fermionic model describes a large set of models that spans from
repulsive systems, $B_{2}(q)>0$, for low values 
of $q$ to attractive ones for large values of $q$. The  cases
$q=1$ and $q=1.96$ certainly describe a free fermionic system 
(at all orders) and
an ideal gas (up to the second virial coefficient) respectively.
This kind of interpolation between boson-like and fermion-like behavior
 is well known in two dimensions for anyonic systems \cite{W,A}. We have
shown that the simplest $SU_{q}(2)$ fermionic system
plays a similar role in three spatial dimensions.


\begin{thebibliography}{99}
\bibitem{Jimbo} See for example:M.Jimbo ed., {\em Yang-Baxter equation
in integrable systems}, Advanced series in Mathematical Physics V.10
(World Scientific,1990).
\bibitem{Wo} S. L. Woronowicz, Publ. RIMS {\bf 23}, 117 (1987).
\bibitem{Manin} Yu I. Manin, Comm. Math. Phys. {\bf 123},163 (1989).
\bibitem{WZ} J. Wess and B. Zumino, Nucl. Phys.B (Proc. Suppl.)
{\bf 18}, 302 (1990).
\bibitem{Wat} U. Carow-Watamura, M. Schlieker
and S. Watamura, Z. Phys. C {\bf 49}, 439 (1991).
 M. R. Ubriaco, Mod. Phys. Lett. A {\bf 8}, 89 (1993).
 S. Shabanov, J. Phys. A {\bf 26}, 2583 (1993).
 A. Lorek and J. Wess, Z. Phys. C {\bf 67}, 671 (1995).
\bibitem{AV} I. Aref'eva and I. Volovich, Phys. Lett {\bf B264}, 62 (1991).
 A. Kempf, J. Math. Phys. {\bf 35}, 4483 (1994).
 T. Brzezinski and S. Majid, Phys. Lett. {\bf B298}, 339 (1993).
 L. Castellani, Mod. Phys. Lett. A {\bf 9}, 2835 (1994).
M. R. Ubriaco, Mod. Phys. Lett. A {\bf 9},1121 (1994).
A. Sudbery, {\em $SU_q(n)$ Gauge Theory}, hep-th/9601033.
\bibitem{I}  S. Iwao, Prog. Theor. Phys.{\bf 83}, 363 (1990).
 D. Bonatsos, E. Argyres and P. Raychev, J. Phys. A {\bf 24},
L403 (1991).
 R. Capps, Prog. Theor. Phys. {\bf 91}, 835 (1994).
\bibitem{Th}Gang Su and Mo-lin Ge, Phys. Lett. A {\bf 173}, 17 (1993).
J.Tuszy\'{n}ski, J. Rubin, J. Meyer and M. Kibler,
Phys. Lett. A {\bf 175}, 173 (1993).  I. Lutzenko and A Zhedanov, Phys. Rev. E {\bf 50},
 97 (1994) . M. Salerno, Phys. Rev. E {\bf 50},  4528 (1994).
 P. Angelopoulou,S Baskoutas,L. de Falco,A. Jannussis,
R. Mignani and A. Sotiropoulou,  J. Phys. A {\bf 27}, L605 (1994). 
 S. Vokos and C. Zachos., Mod. Phys. Lett. A {\bf 9}, 
1 (1994). M. R-Monteiro, I. Roditi and L. Rodrigues,
Mod. Phys. Lett. B {\bf 9}, 607 (1995).
J. Goodison and D. Toms, Phys. Lett. A {\bf 195}, 38 (1994);
Phys. Lett. A {\bf 198},471 (1995).
\bibitem{An} D. Fivel. Phys. Rev. Lett. {\bf 65}, 3361 (1990).
R. Campos, Phys. Lett. A {\bf 184}, 173 (1994). S. Dalton and
A. Inomata, Phys. Lett. A {\bf 199}, 315 (1995).
A. Inomata, Phys. Rev. A {\bf 52}, 932 (1995). 
\bibitem{G}O. W. Greenberg, Phys. Rev. Lett. {\bf 64}, 705 (1990).
\bibitem{NG} See Y. J. Ng, J. Phys. A 23 (1990) 1203 and references
therein.
\bibitem{T} L. A. Takhatajan, Advanced Studies in Pure Mathematics {\bf 19}
(1989) and references therein.
\bibitem{U2} M. R. Ubriaco, {\em Thermodynamics of a free $SU_{q}(2)$
fermionic system}, hep-th/9605179.
\bibitem{W} F. Wilczek, Phys. Rev. Lett. {\bf 48}, 1144 (1982); 
{\bf 49}, 957 (1982).
\bibitem{A} D. Arovas, R. Schrieffer, F. Wilczek and A. Zee, Nucl. Phys. {\bf B251},
117 (1985).
\end{thebibliography}
\end{document}